\documentclass[12pt,english]{iopart}
\usepackage{graphicx}
\usepackage{dcolumn}
\usepackage{babel}
\usepackage{color}
\usepackage{latexsym}
\usepackage[cm]{aeguill}
\begin{document}

\title{A versatile fabrication method for cluster superlattices}
\author{
Alpha T. N'Diaye$^1$\footnote{corresponding author; Email: ndiaye@ph2.uni-koeln.de}, 
Timm Gerber$^1$, 
Carsten Busse$^1$, 
Josef Myslive\v{c}ek$^2$, 
Johann Coraux$^1$\footnote{permanent adress: Institut N\'{e}el/CNRS, B\^{a}t. D, 25 Rue des Martyrs, F-38042 Grenoble Cedex 9, France} 
and Thomas Michely$^1$}
\address{$^1$ {II. Physikalisches Institut}, Universit{\"a}t zu K{\"o}ln, Z{\"u}lpicher Stra{\ss}e 77, 50937 K{\"o}ln, Germany}
\address{$^2$ Charles University, Faculty of Mathematics and Physics, V Hole\v{s}ovi\v{c}k\'{a}ch 2, 180 00 Praha 8, Czech Republic}
\begin{abstract}
On the graphene moiré on Ir(111) a variety of highly perfect cluster superlattices can be grown as shown is for Ir, Pt, W, and Re.  Even materials that do not form cluster superlattices upon room temperature deposition may be grown into such by low temperature deposition or the application of cluster seeding through Ir as shown for Au, AuIr, FeIr.  Criteria for the suitability of a material to form a superlattice are given and largely confirmed. It is proven that at least Pt and Ir even form epitaxial cluster superlattices. The temperature stability of the cluster superlattices is investigated and understood on the basis of positional fluctuations of the clusters around their sites of minimum potential energy. The binding sites of Ir, Pt, W and Re cluster superlattices are determined and the ability to cover samples macroscopically with a variety of superlattices is demonstrated. 
\end{abstract}

\maketitle

\section{Introduction}

Clusters are a distinct state of matter. Not only their structure, electronic, magnetic and optical
properties change with size, but also new properties emerge as a result of quantization
effects unknown in the atom and the bulk solid \cite{khanna2003}. Well known examples for unique cluster properties
are magic sizes in geometric and electronic structure, superparamagnetism, plasmon resonances or
size dependent reactivity \cite{haberland2005,kreibig1995,heiz2006}. Though cluster properties may be studied in their purest form for mass selected clusters in the free beam, a use of clusters requires a suitable support \cite{meiwes2000}. Ideally, a system of supported clusters is a regular array with equal sized, equally spaced clusters, each in an identical
environment and of macroscopic array extension. The cluster bonding to the substrate should be
strong enough to warrant their stability at the temperature of operation and weak enough not to
destroy them as own entities. The substrate should be inert in a sense that it does not deteriorate
under the condition of use but active to bestow new functionalities to the clusters without destroying
them. Such arrays or cluster superlattices would allow one to address single clusters (as
might e.g. be of interest for magnetic applications), to obtain a large amplitude of response resulting
from the additive superposition of all single clusters and being characteristic for a cluster of well
defined size in a specific environment (as e.g. needed in catalysis), or to obtain a collective coherent
response characteristic of a large interacting ensemble of clusters (as e.g. in optics).

This vision triggered attempts to make use of self organization and/or templates for the creation of
cluster superlattices and a few examples are given here (compare also \cite{becker2008}). By making use of the regular arrangement of steps on vicinal Au(111) and its herringbone reconstruction crossing the steps at right angles a proper Co cluster superlattice could be realized \cite{repain2002}, enabling the measurement of its temperature dependent magnetic properties including the distribution of cluster magnetic anisotropy energies \cite{weiss2005}. Also Fe could be grown regularly \cite{rohart2008}, but the method is limited by its low growth temperature and strong cluster substrate interaction giving rise to alloying. Alumina double layers on Ni$_{\rm{3}}$Al(111) exhibit a phase with a regular superlattice of sites for nucleation upon metal deposition \cite{degen2004}, which are holes in the oxide \cite{schmid2007}. Cluster superlattices were realized for a variety of metals at room temperature \cite{degen2004, hamm2006,lehnert2006} and a high perfection was achieved for Pd \cite{degen2004}. Though such cluster lattices appear to be suited as model catalysts, the preparation of the alumina layer is the result of a subtle procedure, comprises defects and at least two phases of alumina with limited domain sizes, of which only one is suitable as a template \cite{schmid2007}.
Recently Xe buffer layer assisted self assembly of Co on a BN-layer forming a moiré with Rh(111) was achieved \cite{zhang2008}. Though the placement of the Co clusters is regular, the method hampers from the very low
temperature of fabrication, the limited thermal stability and the only incomplete filling of the template cells.
As a final example, on the W(110)/C-R(15$\times$12) surface carbide arrangements of Au, Ag and Co clusters could be grown,  the latter two with a high degree of order \cite{varykhalov2008,bachmann2008}. Though the thermal stability of the arrangements is rather good, no macroscopic extension of a superlattice is achieved due to the persistent existence of terraces without the surface carbide. 

Graphene moirés with noble metal surfaces are a new and unique support for cluster superlattices.
This was shown first for graphene moirés on Ir(111) allowing Ir cluster superlattice formation \cite{ndiaye2006}.
Though a special material system, the binding mechanism is versatile. Through metal deposition on the graphene layer
-- which is only weakly interacting with the metal substrate in the absence of metals deposits \cite{pletikosic2009} -- graphene locally rehybridizes from sp$^2$ to sp$^3$ carbon bonds at a specific location in the moiré unit cell and in between the substrate and deposit metal, thereby forming strong carbon metal bonds \cite{feibelman2008}. In between the substrate and deposit metal the sp$^3$ rehybridized graphene has tetrahedral bond angles and may thus be considered as diamondlike. Within each moiré unit cell rehybridization is possible where locally the carbon rings center around a threefold coordinated hcp-site or a threefold coordinated fcc-site (both having three C-atoms on atop substrate sites). These locations are named hcp region or fcc region, respectively.  Experimentally and by DFT calculations we find that Ir clusters bind by far stronger to hcp regions \cite{ndiaye2006,feibelman2008,ndiaye2008}.

Fig. 1a displays a cluster superlattice grown on a graphene flake formed by temperature programmed growth on Ir(111) \cite{coraux2009}. With this method graphene covers the surface only partially, which is suitable for scanning probe investigations, as it retains the metal surface partially for calibration of cluster size through island coverage in the graphene free areas. Also direct comparison of the properties of clusters and those of the deposit islands on the metal is possible. Recently, we employed chemical vapor deposition at high temperatures to grow graphene fully covering the substrate and displaying hardly any defects \cite{coraux2009,coraux2008}. Using a dedicated combination of both methods \cite{vangastel2009}, we are not only able to ensure full coverage but also the graphene and Ir dense packed atomic rows to be parallel, i.e. no traces of rotational variants are present \cite{loginova2009}.

In this manuscript we will show that graphene moirés as active templates are superior to 
other systems enabling the growth of two dimensional cluster superlattices on solid surfaces through 
a unique combination of properties: (i) the cluster binding mechanism is universal enabling growth of a large diversity of materials as superlattice; (ii) the superlattice order is extremely high with a completely filled lattice;(iii) if desired the superlattice extends macroscopically without uncovered substrate patches; (iv) fabrication is possible at room temperature; (v) the superlattices possess a reasonably high thermal stability and display absence of alloying and interdiffusion within a large temperature range; (vi) the cluster size is tunable and the size distribution is narrow.

To demonstrate the universality of our approach in view of superlattice forming materials we primarily focussed on cluster materials with potentially interesting structural, magnetic, catalytic or optical properties. As heuristic guidelines for the suitability of a material to form a superlattice we considered three factors: (i) A large cohesive strength of the material as an indicator for the ability to form strong bonds. (ii) A large extension of a localized valence orbital of the deposit material allows it to efficiently interact with the graphene $\pi$-bond and thus to initiate rehybridization to diamondlike carbon underneath the cluster. (iii) A certain match of the graphene unit cell repeat distance on Ir(111) of 2.452\,\AA~\cite{ndiaye2008} and the nearest neighbour distance of the deposit material is necessary to fit the first layer cluster atoms atop of every second C atom. As small clusters - which are the relevant sizes to start superlattice growth - have a smaller lattice parameter compared to bulk materials, and as Ir works perfectly as a cluster material \cite{ndiaye2006}, we considered 2.7\,\AA~as an optimal nearest neighbor distance\cite{footnote}. Table compares the data for the tested materials.
Tungsten was selected as a likely candidate for superlattice formation (all three figures are in favor of superlattice formation), but with a different crystal structure than Ir. Re, Pt and Au were selected as as potentially interesting materials for catalysis. According to our guidelines we expected Re to be most likely a superlattice forming material, also Pt, but with a slightly smaller probability and Au as an unlikely candidate. Fe and Ni were selected as materials because of their ferromagnetism. According to their figure of merits, however, we did not expect superlattice formation.

\begin{table}
\begin{center}
\begin{tabular}{|c|c|c|c|c|}
\hline
material & structure & $a$/\AA & $r_{\rm d}$/pm & $E_{\rm{coh}}$/eV \\\hline\hline
Ir & fcc & 2.715 & 70.8& 6.94  \\\hline
W & bcc & 2.741 & 77.6 & 8.90 \\\hline
Re & hcp & 2.761 & 73.9 & 8.03 \\\hline
Pt & fcc & 2.775 & 65.9 & 5.84 \\\hline
Au & fcc & 2.884 & 63.5 & 3.81 \\\hline
Fe & bcc & 2.483 & 38.2 & 4.28 \\\hline
Ni & fcc & 2.492 & 33.8 & 4.44 \\\hline
\end{tabular}
\end{center}
\caption{Crystal structure, nearest neighbor distance $a$, valence d-orbital radius $r_{\rm d}$ and cohesive energy $E_{\rm{coh}}$ of the tested materials for cluster superlattice formation on the graphene moiré on Ir(111).
}
\label{tabelle}
\end{table}

\section{Experimental}
Experiments were carried out in an ultra high vacuum variable temperature scanning
tunneling microscopy apparatus with a base pressure of $1 \times 10^{-10}$\,mbar 
equipped with a mass separated ion source, low energy electron diffraction, a gas
inlet for ethylene exposure through a tube ending approximately 2\,cm above
the sample surface and a four pocket e-beam evaporator.
The sample was cleaned by repeated cycles of sputtering with a beam of 2.5\,keV
Xe$^+$ ions at 300\,K and annealing to 1520\,K. Graphene was prepared using
temperature programmed growth (TPG) at 1470\,K or chemical vapor deposition (CVD) at 1370\,K
\cite{coraux2009, vangastel2009}. TPG results in a graphene surface coverage of about 25\% with
flakes of a few 100\,\AA~to about 1000\,\AA~extension. Adding a CVD step results in up to full
graphene surface coverage and domains of $\mu$m and larger extension. Cluster growth was
performed through e-beam evaporation of well degased high purity metals with typical
deposition rates of a few times 10$^{-2}$\,ML/s where 1\,ML (one monolayer)
corresponds to the surface atomic density of Ir(111). During deposition the pressure
remained in the low 10$^{-10}$\,mbar range. Precise calibration of the deposited
amount $\Theta$ was obtained by STM image analysis of the fractional area of deposit islands
on the clean Ir(111) surface after a defined deposition time. If necessary the sample 
was gently annealed prior to coverage calibration. Through annealing the heteroepitaxial islands became compact and their size increased, thereby minimizing imaging errors associated with the finite STM tip size. 
Clusters were imaged at deposition temperature, if not indicated otherwise. However, 
for deposition above 300\,K or after annealing the sample was quenched to 300\,K for imaging.  

\section{Results}
\subsection{Cluster structure and superlattice formation at 300\,K}
\begin{figure}
\begin{center}
\includegraphics{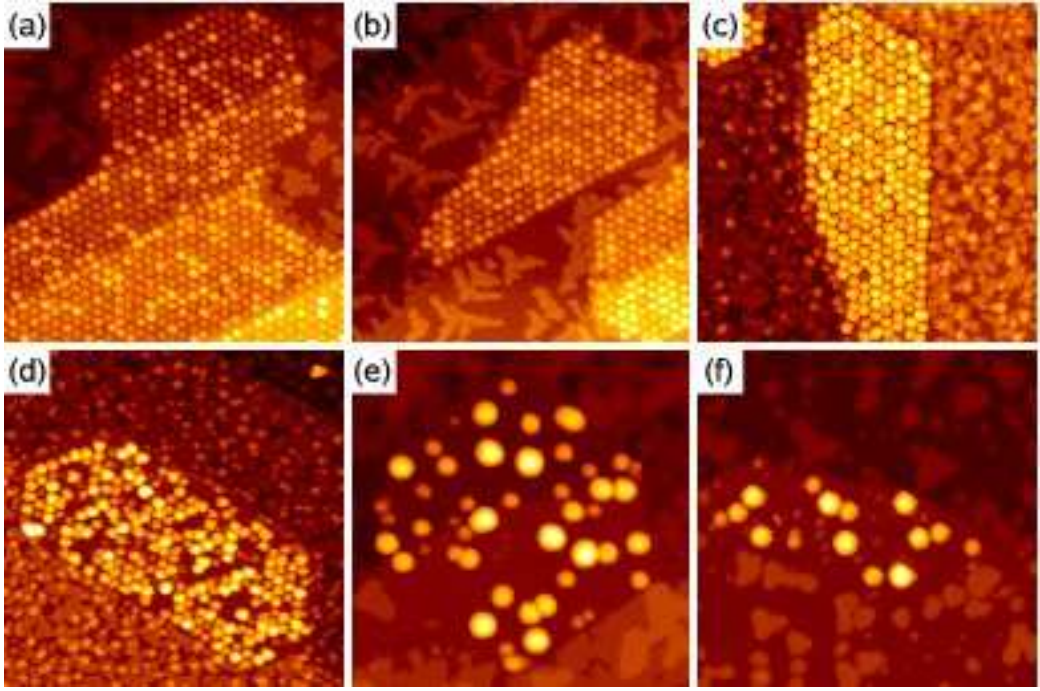}
\end{center}
\caption{STM topopgraphs of Ir(111) with graphene flakes after deposition of an amount $\Theta$ of various metals at 300\,K.
(a) $\Theta = 0.20$\,ML Ir, average cluster size $s_{\rm{av}} = 17$\,atoms;
(b) $\Theta = 0.25$\,ML Pt, $s_{\rm{av}} = 22$\,atoms;
(c) $\Theta = 0.44$\,ML W,  $s_{\rm{av}} = 38$\,atoms;
(d) $\Theta = 0.53$\,ML Re,  $s_{\rm{av}} = 60$\,atoms;
(e) $\Theta = 0.77$\,ML Fe,  $s_{\rm{av}} = 420$\,atoms;
(f) $\Theta = 0.25$\,ML Au,  $s_{\rm{av}} = 100$\,atoms.
Image size 700\,\AA~$\times$~700\,\AA.}
\label{compare300K}
\end{figure}

Figure 1 displays STM topographs after deposition of 0.2-0.8\,ML on Ir(111) partly covered with graphene flakes. The graphene flakes are typically attached to substrate steps, but also extend over one or several of them [Fig. 1(a), 1(b), 1(e)]. In the areas without graphene, deposit metal islands of monolayer height formed from the evaporated material. In Figs. 1(c)-1(e) already second layer island nucleation took place. Depending on the deposited metal the island nucleation density on Ir(111) varies considerably being highest for the W and Re, the metals with the highest cohesive energy [Figs. 1(c) and (d)]. The deposit islands mostly reflect the threefold symmetry of the substrate. It is obvious from Fig. 1 that all deposited materials are pinned to graphene flakes to a certain extend and that graphene on Ir(111) is in all cases much more sticky to the deposited metals than the surface of graphite \cite{bardotti1995}. However, not all materials form a cluster superlattice. 

Ir and Pt form superlattices of similar perfection [compare Figs. 1(a) and 1(b)]. For the represented $\Theta \approx 0.2$\,ML both materials exhibit two distinct height levels of the clusters indicating an out-of-plane texture of the cluster orientation. Distinct height levels are present also for larger and higher clusters up to the coalescence threshold (compare also Fig. 1 of reference \cite{ndiaye2006}). The apparent height differences between 3\,ML and 4\,ML clusters as well as those between 4\,ML and 5\,ML clusters are for Ir and Pt in the range of 2.2\,\AA~and 2.3\,\AA, i.e. of the size of a monatomic step height $h_{\rm 1}$ on a (111) terrace [$h_{\rm{1,Ir}} = 2.22$\,\AA~and $h_{\rm{1,Pt}} = 2.27$\,\AA]. The height differences in lower levels differ from these numbers. Specifically, the apparent height difference between the graphene substrate and the first cluster height level for $s_{\rm{av}} > 15$ atoms is always found to be larger than 2.5\,\AA~while the difference between the first and second height level is typically below 2.0\,\AA. We interpret these deviations from the (111) step height as density of state effects. While for Ir and Pt the height of each single cluster is an integer number of (111) layers, the average cluster height $h_{\rm{av}}$ may be noninteger due to the averaging. Fig. 2 displays an analysis of $h_{\rm{av}}$ for Ir clusters grown at 350\,K and Pt clusters grown at 300\,K. The decrease of $h_{\rm{av}}$ for Ir clusters between 1.5\,ML and 2.0\,ML is due to the onset of cluster sintering. Upon cluster sintering the clusters reshape and material flows into the gaps separating the clusters, thereby causing frequently a height reduction. It is also apparent from Fig. 2, that large Pt clusters tend to grow flatter than Ir clusters, giving rise to a somewhat earlier cluster sintering. 
\begin{figure}
\begin{center}
\includegraphics[width=0.6 \linewidth]{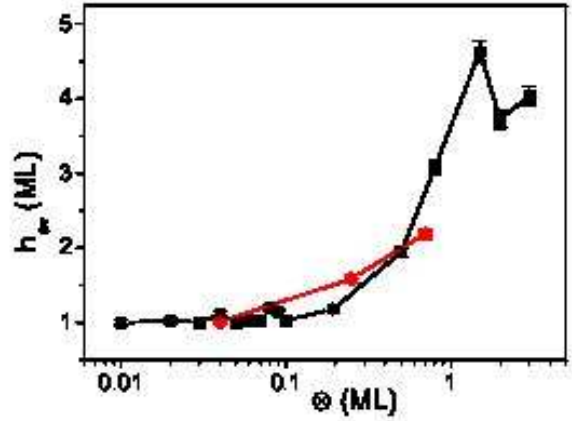}
\end{center}
\caption{Ir (full squares) and Pt (red dots) average cluster heights $h_{rm{av}}$ in monolayers as a function of deposited amount $\Theta$.}
\label{heights}
\end{figure}

After having established the $\left[111\right]$ out-of-plane texture of Pt and Ir clusters, the question arises whether the clusters possess also in-plane-texture, i.e. whether the clusters are also oriented within the surface plane. Fig. 3(a) displays a small area of an Ir cluster array after deposition of 1.5\,ML imaged with a large tunneling resistance of $2.3 \times 10^9\,\rm{\Omega}$. Despite the absence of atomic resolution at least some cluster edges appear to be oriented along the $\left\langle 211\right\rangle$-direction. Lowering the tunneling resistance to $7 \times 10^6\,\rm{\Omega}$ - thereby bringing the tip close to the cluster surfaces - and tuning the contrast to the different levels of the cluster top mesas changes the picture. The top mesas are almost atomically resolved (some row like corrugation is visible) and the edges of the top mesas are unambiguously oriented along $\left\langle 1\bar{1}0\right\rangle$. It needs to be noted that under low tunneling resistance conditions necessary for atomic resolution clusters are usually picked up by the STM tip. The apparently different orientation of cluster edges at high tunneling resistances is a mere imaging artifact.
If a tip scans a hexagonal grid of elevated objects in a large distance not the shape of the objects forming the grid but the hexagonal grid itself determines the apparent orientation of the boundaries between the objects. The inset of Fig. 3(d) represents a ball model of a five layer cluster containing 140 atoms which is consistent with the experiments. Although we are unable to prove directly that the cluster sidewalls are formed by $\left\{100\right\}$ and $\left\{111\right\}$ facets, facets of smaller slope would imply cluster contacts at their base. This is unlikely to be the case, as we find in annealing sequences clusters to reshape rapidly upon contact.
To summarize, we have shown that Ir clusters (and the Pt ones most likely as well) are \emph{epitaxial} clusters with the (111) cluster planes parallel to the substrate surface and the $\left\langle 1\bar{1}0\right\rangle$ cluster directions parallel to the $\left\langle 1\bar{1}0 \right\rangle$-directions of the Ir substrate and the $\left\langle1\bar{1}20\right\rangle$-directions of graphene. This epitaxy of Ir clusters on the graphene moiré is also predicted by the geometry in the DFT based model of cluster binding \cite{feibelman2008}.

\begin{figure}
\begin{center} 
\includegraphics{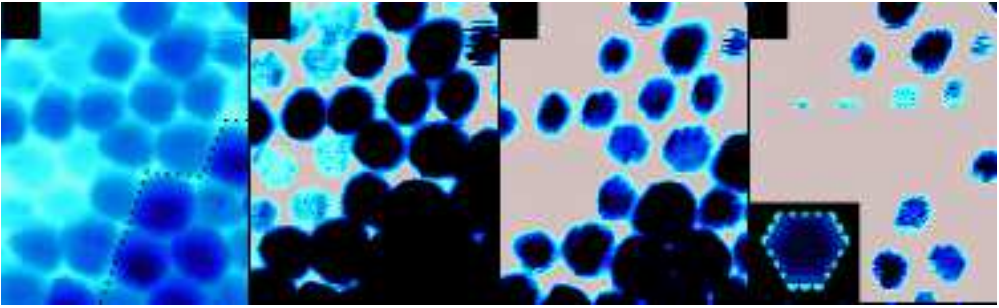}
\end{center}
\caption{
(a) STM topograph after deposition of 1.5\,ML Ir on a graphene moiré on Ir(111), $s_{av} = 130$. The dashed line indicates the position of an underlying substrate step. The visible clusters on the lower terrace have heights of 4\,ML and 5\,ML.  For imaging the tunneling resistance was $2.3 \times 10^9\,\rm{\Omega}$ ($I = 0.5$\,nA and $U = 1.15$\,V. (b),(c),(d) Same location imaged successively with a low tunneling resistance of $7 \times 10^6\,\rm{\Omega}$ ($I = 30$\,nA and $U = 0.2$\,V) and contrast tuned to the different cluster levels. Cluster edges are aligned along $\left\langle 1\bar{1}0\right\rangle$ substrate directions [see (b)]. Inset in (d): Ball model of a 5 layer cluster of 140 atoms consistent with the experimental observations. Image size 130\,\AA~$\times$~160\,\AA.}
\label{clusteraligned}
\end{figure} 

Also tungsten forms a cluster superlattice of high perfection. Compared to Pt and Ir the centers of mass of the clusters deviate slightly more from a perfect hexagonal superlattice. Still the scatter of the cluster positions is too small to result in cluster sintering for $\Theta = 0.44$\,ML as visualized in Fig. 1c. Close inspection of our data - also for $\Theta = 0.04$\,ML - shows that less than 1\% of clusters is out registry with the hcp regions (see also below).
The apparent cluster heights for $\Theta = 0.44$\,ML range from 4\,\AA~to 8\,\AA~ with an average around 6\,\AA. The clusters seem higher than the Ir ones for a comparable $\Theta$. Distinct height levels are also present for W clusters as is obvious from Fig. 1(c). However, the height levels are less well defined due to their smaller separation. Between adjacent clusters we measure frequently height differences of about 0.8\,\AA~to 1\,\AA. With a small distortion the dense packed W(110) plane with a nearest neighbour distance of 2.74\,\AA~would fit to the triangular 2.46\,\AA~periodicity provided by the graphene for rehybridization. However, a $\left[110\right]$ out-of-plane cluster texture would result in height levels spaced by 2.24\,\AA, i.e. a similar spacing as for Ir and Pt. From the presence of intermediate levels we exclude this cluster orientation as the single one. Small step heights of 0.8\,\AA~to 1\,\AA~are consistent with a $\left[111\right]$ out-of-plane texture with a step spacing of 0.91\,\AA. The open W(111) plane has also the threefold symmetry of the graphene moiré in the hcp or fcc regions. However, in plane atomic spacings are 4.48\,\AA. They could be fitted to the 4.27\,\AA~ separation of a $(\sqrt{3} \times \sqrt{3}) R 30$ superstructure of graphene. For a final statement more detailed experiments are certainly necessary but also likely to be rewarding to settle the issue of the W cluster superlattice texture.

For Re only partial order of the cluster arrangement is realized as visible in Fig. 1(d). Two possible reasons could cause the apparent disorder: (i) the existence of several adsorption mimima within a moiré unit cell, causing less ordered growth followed by early coalescence and cluster rearrangement; (ii) a too low depth of the cluster size dependent adsorption mimina. Assuming an increase of the potential well with increasing cluster size, adatoms and small clusters up to a certain size $s_{\rm c}$ would be likely to leave their unit cell during growth. Both scenarios would give rise to the observed heterogeneous size distribution, but the latter would explain also the existence of a large number of empty moiré unit cells. For Re we performed also low coverage experiments [compare Fig. 7(d)]. We find a comparatively low moiré unit cell occupation probability (or a comparatively large $s_{\rm{av}}$) for the deposited amount, but with few exceptions all clusters adsorbed on a regular grid. We are therefore convinced that indeed the adsorption site minima are not of sufficient depth to keep small Re clusters up to a certain critical size (being much larger than the one for Ir, Pt and W) during and after growth on their position within the moiré. We note that the poor quality of the Re cluster superlattice is at variance with our expectations, as Re has a large cohesive energy, an extended d-orbital and a reasonably well matching nearest neighbour distance (compare table 1).

For Fe [Fig. 1(e)], Au [Fig. 1(f)] and Ni (data not shown) at room temperature no superlattice can be realized. Large clusters, lacking defined height levels or shape features are formed. The unstructured, hemispherical clusters display  heights of up to 23\,\AA~for Fe and 15\,\AA~for Au. The absence of a regular cluster superlattice for these materials is expected in view of their small cohesive energy and/or their limited valence orbital extension. Binding of the deposit metal to graphene is apparently too weak to trap adatoms and small clusters, i.e. the depth of the potential energy minima within a moiré unit cell is not sufficient to stabilize a growing cluster.   

\subsection{Low temperature cluster superlattice growth and annealing}

\begin{figure}
\begin{center} 
\includegraphics{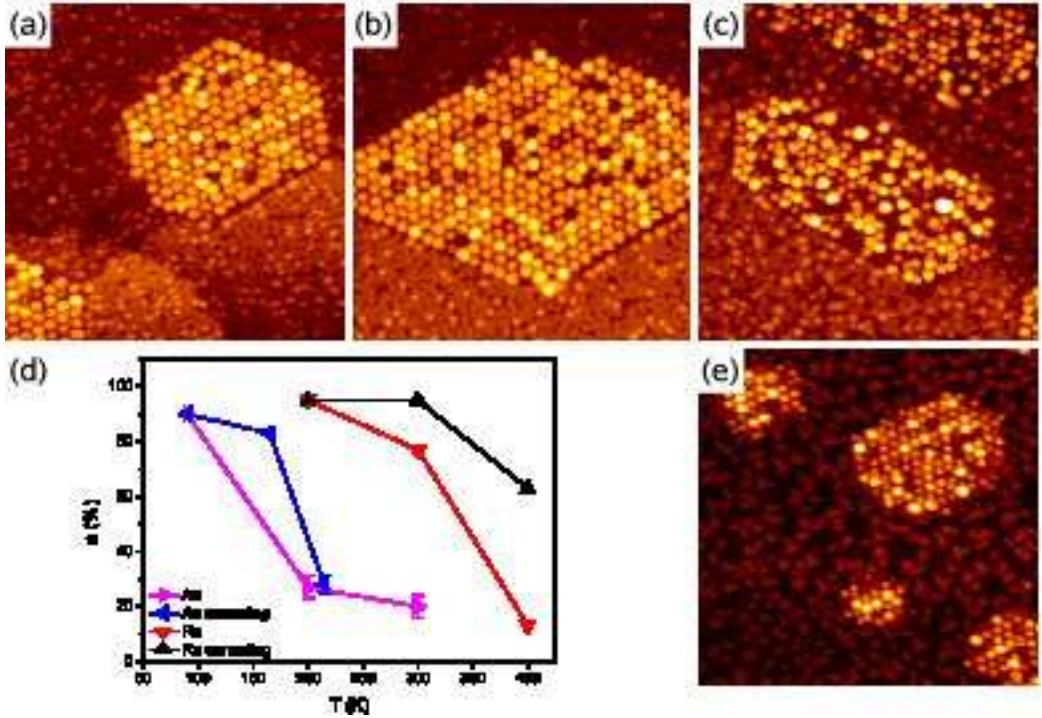}
\end{center}
\caption{
STM topographs of Ir(111) with graphene flakes after depositing amounts $\Theta$ of Re or Au at the indicated temperature $T$.
(a) $\Theta = 0.45$\,ML Re, $s_{\rm{av}} = 41$, $T = 200$\,K;
(b) $\Theta = 0.45$\,ML Re, $s_{\rm{av}} = 41$, $T = 200$\,K. Subsequently the sample has been annealed to 300\,K and imaged; 
(c) $\Theta = 0.53$\,ML Re,  $s_{\rm{av}} = 60$, $T = 300$\,K;
(d) Occupation probability $n$ of moiré unit cells with clusters as a function of growth and annealing temperature $T$; triangles pointing to the right: Au clusters, 0.25\,ML deposited at the indicated $T$; triangles pointing to the left: Au clusters, 0.25 ML deposited at 90\,K and additionally annealed to the indicated $T$; down triangles: Re clusters, 0.45\,ML deposited at the indicated $T$; up triangles: Re clusters, 0.45\,ML deposited at 200\,K and annealed to the indicated $T$.
(e) $\Theta = 0.25$\,ML Au,  $s_{\rm{av}} = 24$, $T = 90$\,K.}
\label{anneal1}
\end{figure} 

If we assume that cluster superlattice formation for Re and even more for Au, Fe and Ni is impeded by a too high mobility of small clusters during growth, lowering the growth temperature would decrease $s_{\rm c}$ and thus be an efficient strategy to improve superlattice formation. Fig. 4(a) displays a Re cluster superlattice grown at 200\,K by deposition of $\Theta = 0.45$\,ML. Annealing to 300\,K does not change the cluster superlattice as visible in Fig. 4(b). The side-by-side comparison to Re clusters grown at 300\,K in Fig. 4(c) makes the efficiency of lowering the growth temperature to improve superlattice formation obvious. The low growth temperature reduced $s_{\rm c}$, enabled the almost complete filling of the moiré unit cells and allowed the clusters to grow to a size being stable even at 300\,K. One might speculate that a further lowering of the growth temperature would have improved the Re cluster superlattice even further. In the quantitative analysis of Fig. 4(d) it is also apparent that the difference in moiré unit cell occupation between growth at 400\,K and annealing to 400\,K from a formation temperature of 200\,K is even larger than the corresponding one at 300\,K. At the same time it is also obvious from the data of Fig. 4(d) that growth at a low temperature and annealing preserves the superlattice only up to a limited temperature [(compare up and down triangles at 400\,K in Fig. 4(d)]. Lowering the growth temperature to 200\,K for Au did not significantly enhance cluster nucleation. However, lowering the growth temperature even further to 90\,K resulted in an Au cluster superlattice of moderate order as shown in Fig. 4(e). Also in this case the once formed superlattice may be preserved to a higher temperature. However, as visible from the quantitative analysis shown in Fig. 4(d), the Au cluster superlattice is deteriorated already after annealing to 220\,K. 

Mild annealing leads to more subtle effects which do not affect the positional order of the cluster array. For 0.45 ML of Ir deposited at 300\,K, by annealing to 450\,K for 300\,s the amount of single layered clusters decreases from $(15\pm2)$\,\% to $(10\pm2)$\,\% with a negligible decrease of the overall occupational density of the moiré with clusters. 
This implies, that single layered clusters of a certain size transform to a more stable two layered form upon annealing.

\subsection{Cluster seeding}

\begin{figure}
\begin{center} 
\includegraphics{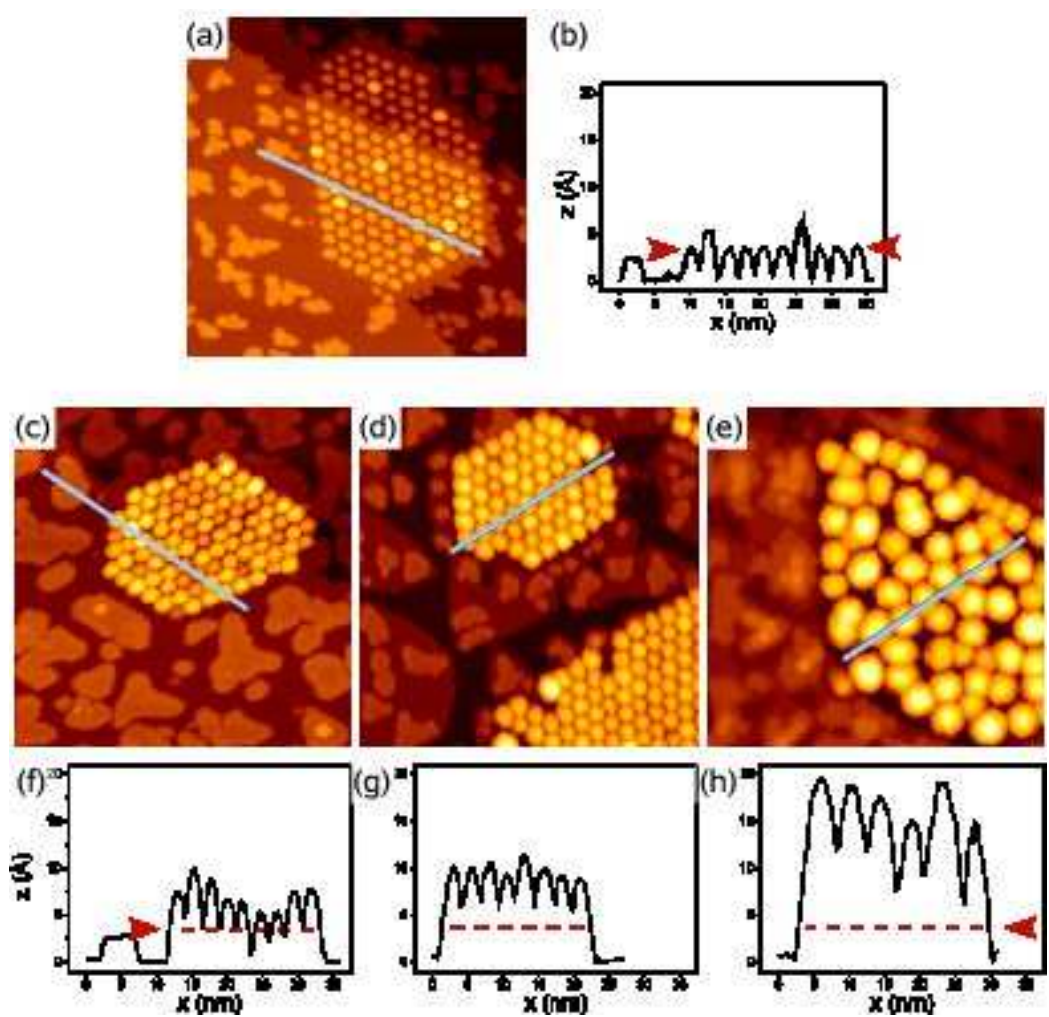}
\end{center}
\caption{
(a), (c), (d),(e) STM topographs of Ir(111) with graphene flakes after depositing amounts $\Theta$ of Ir and Au or Fe at
300\,K. Image size 500\,\AA~$\times$~500\,\AA.
(a)~$\Theta = 0.10$\,ML Ir, $s_{\rm{av}} = 9$. 
(b)~Line scan along the line indicated in (a). The two red arrows indicate the height level of ML clusters.
(c)~$\Theta = 0.10$\,ML Ir and subsequently $\Theta = 0.35$\,ML Au corresponding to 30 Au atoms per seed cluster.
(d)~$\Theta = 0.10$\,ML Ir and subsequently $\Theta = 0.70$\,ML Fe corresponding to 61 Fe atoms per seed cluster.
(e)~$\Theta = 0.10$\,ML Ir and subsequently $\Theta = 2.0$\,ML Fe. Average number of Fe atoms per cluster is 760.
(f)~Line scan along the line indicated in (d). 
(g)~Line scan along the line indicated in (e).
(h)~Line scan along the line indicated in (f). 
The red dashed lines in (f)-(h) indicate the height level of monolayer Ir clusters in (b).}
\label{seeding}
\end{figure} 

Low cohesive energy metals tend not to form cluster superlattices on the graphene moiré on Ir(111), as tested by us for Au, Fe, and Ni. Such metals wet high cohesive energy metals, due to their lower surface free energy. High cohesive energy metals mostly form rather perfect cluster superlattices, tested here for Ir, Pt and W. It is thus natural to apply cluster seeding, i.e. to define the positions of the clusters by a small $\Theta$ of a high cohesive energy metal and to grow these seeds by subsequent deposition of a low cohesive energy metal \cite{schmid2007,hamm2006}. The successful application of this method is visualized in Fig. 5.  As shown in Fig. 5(a) first through deposition of 0.1\,ML Ir seed clusters are created in nearly all moiré unit cells. Subsequent deposition of Au [Fig. 5(b)] or Fe [Fig. 5(c)] at 300\,K results in highly perfect Au and Fe cluster superlattices with Ir cores. Comparing Fig. 1(f) with Fig. 5(c) for Au and of Fig. 1(e) with Fig. 5(d) for Fe makes the dramatic effect of seeding obvious. Even after deposition of amounts beyond the coalescence threshold the seeding has beneficial effects on cluster uniformity and distribution. For deposition of $\Theta = 2$\,ML Fe still a high density of large, uniformly sized and spaced Fe clusters grows. These Fe clusters contain 760 Fe atoms and have a height of 20\,\AA~[compare Fig. 5(h)]. They are not more positioned on a superlattice, but span  4 to 5 moiré unit cells and contain about 40 Ir atoms.  

\subsection{Temperature stability}

\begin{figure}
\begin{center} 
\includegraphics{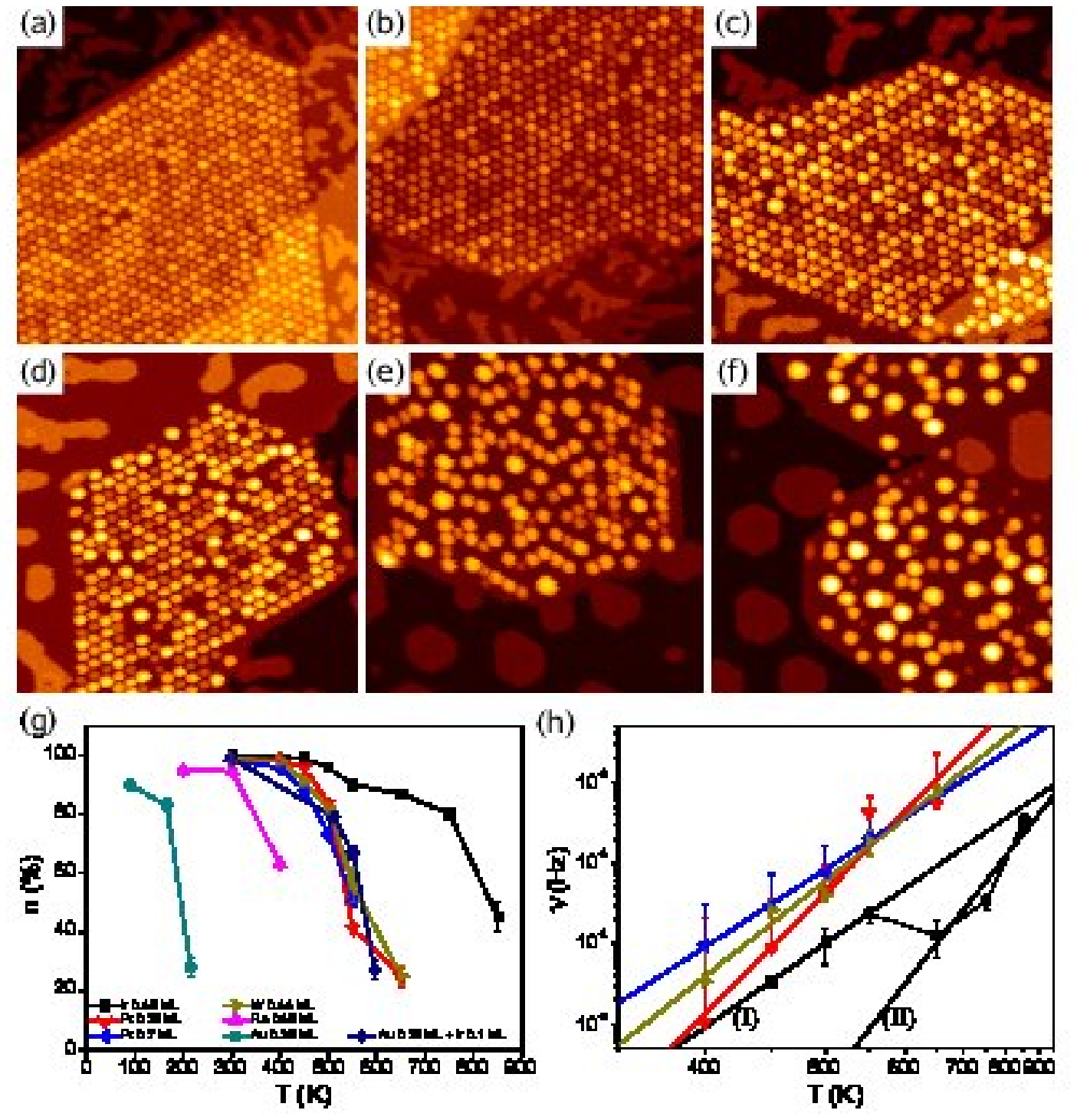}
\end{center}
\caption{(a)-(f) Annealing sequence of a Pt cluster superlattice on a graphene moiré on Ir(111) with $\Theta = 0.25$\,ML grown at (a) 300\,K and subsequently annealed in 300\,s time intervals to (b) 400\,K, (c) 450\,K (d) 500\,K, (e) 550\,K and (f) 650\,K. Image size 700\,\AA~$\times$~700\,\AA.
(g) Occupation probability $n$ of moiré unit cells with clusters as a function of annealing temperature $T$.
(h) Arrhenius plot of cluster hopping rate $\nu(T)$. Lines represent fits for the hopping rate with diffusion parameters as shown in table \ref{tab:diffusionparams}. For Ir, two parts of the dataset (I) and (II) are fitted independently.
}
\label{anneal2}
\end{figure} 

For applications of cluster superlattices in nanomagnetism and nanocatalysis thermal stability of the cluster arrays and the absence of sintering at the temperature of use is of crucial importance. To provide data in this respect we investigated the thermal stability of the materials tested so far and display the results in Fig. 6. The example annealing sequence of Figs. 6(a) to 6(f) shows the gradual decay of the cluster superlattice through isochronal annealing steps of 300\,s up to 650\,K. The Pt cluster superlattice remains intact up to 400\,K. Fig. \ref{anneal2}(g) quantifies annealing by plotting the temperature dependence of the moiré unit cell occupation probability $n$ as a function of temperature $T$. It is apparent that the Ir cluster superlattice is indeed the most stable one, decaying as the only one in two steps. Most cluster superlattices are stable up to 400\,K, which provides a reasonable temperature window for nanocatalysis and nanomagnetism experiments. 

\begin{figure}
\begin{center} 
\includegraphics{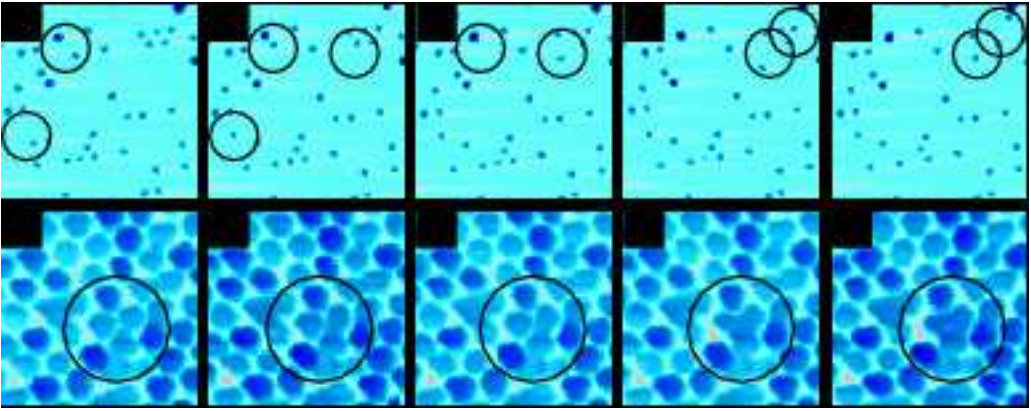}
\end{center}
\caption{(a)-(e) STM topographs after deposition of 0.01\,ML Ir at 350\,K and subsequent heating to 390\,K, The time lapse sequence images the same surface spot every 120\,s at 390\,K. Image size 250\,\AA~$\times$~250\,\AA, $s_{\rm{av}} = 4.5$ atoms. Circles indicate, where changes take place in successive images.
(f)-(j) STM topographs after deposition of 2.0\,ML Ir at 350\,K and subsequent heating to 470\,K, The time lapse sequence images the same surface spot every 120\,s at 470\,K. Image size 150\,\AA~$\times$~150\,\AA.}
\label{dynamics}
\end{figure} 

The decay of all cluster superlattices occurs due to the thermally activated motion of clusters. The clusters fluctuate around their equilibrium positions within the moiré unit cell. The magnitude of cluster fluctuations depends on the cluster size and the internal cluster structure (the isomer). Upon encounter during their fluctuations the clusters merge immediately, on a time scale of less than a second. The outcome of the merging again depends on cluster size. We distinguish two prototypical situations. (i) The clusters consist only out of a few atoms. Such clusters result from a very low deposited amount $\Theta$. 
For such a $\Theta$ also the moiré unit cell occupation probability $n$ is typically well below 1. 
Through thermally activated fluctuations a cluster may surmount the activation barrier $E_a$ to leave its moiré unit cell. If the cluster arrives in an empty cell it will rest there, if it arrives in an occupied cell the two clusters in the cell coalesce. They reshape completely such that as end product a single, compact cluster results which is located entirely within a single moiré unit cell. This is the regime of \emph{complete} cluster coalescence. Figs. 7(a)-(e) display a sequence of STM topographs taken at 390\,K. The visible clusters result from $\Theta = 0.01$\,ML and have $s_{\rm{av}} = 4.5$ atoms. White circles in subsequent stills indicate locations of thermally activated changes. The white circles in the upper left corner of Figs. 7(a)-(c) highlight a situation of thermally activated cluster motion resulting eventually in complete cluster coalescence. Note that the resulting cluster appears to be larger and higher in consequence of complete coalescence. In the Pt annealing sequence of Fig. 6 in (d) a considerable number of larger and higher clusters appear, which are preferentially located next to empty moiré cells. They are likely to be formed by complete coalescence. In addition, in Figs. 7(a)-(e) a number of thermally activated cluster jumps into empty cells are circled, which consequently do not result in cluster coalescence. (ii) If the clusters are large, close to the coalescence threshold, cluster merging proceeds differently. Still such clusters fluctuate around the location of their potential energy minimum. According to their large mass and better internal stability the magnitude of fluctuations has diminished and one might ask how cluster merging takes place at all. However, only small fluctuations are necessary to initiate cluster merging, as due to their large size the cluster have only a small edge separation of a few \AA ngstr\"{o}ms. For coalescence the clusters do not have to leave their moiré unit cell, but it is sufficient to move a little up their shallow potential energy depression to encounter a neighbouring cluster. Due to their size and significant binding the resulting cluster spans two moiré unit cells and does not reshape completely. This is the regime of cluster \emph{sintering} or \emph{incomplete} cluster coalescence. Figs. \ref{dynamics} (f)-(j) display a situation of cluster sintering imaged at 470\,K for clusters formed after deposition of 1.5\,ML Ir (130 atoms per moiré unit cell) at 350\,K. Two subsequent cluster sintering events of neighbouring clusters result eventually in a new single cluster extending over three moiré unit cells. The two scenarios depicted above are extreme cases and intermediate situations occur. From what has been said above it appears that cluster superlattice stability depends also on $s_{\rm{av}}$. It is expected that arrays of medium sized clusters with diminished fluctuation amplitudes and still sufficient separation from their neighbours are the most stable ones.

To obtain a quantitative estimate for parameters determining cluster superlattice decay we model it as follows.
We assume the cluster superlattice to consist of clusters with a unique activation energy $E_{\rm a}$ for cluster interaction with a neighbouring cluster and an interaction frequency $\nu = \nu_{\rm 0} e^{-E_{\rm a}/k_{\rm B} T}$, where $\nu_{\rm 0}$ is an attempt frequency characterizing the frequency of cluster fluctuations, $k_{\rm B}$ the Boltzmann constant and $T$ the temperature of the experiment.
The probability that one cluster encounters another one is proportional to $n$. We assume complete coalescence, i.e. the final cluster to occupy only a single moiré unit cell. The number of these events is as well proportional to $n$. Under these conditions the decrease of $n$ with time $t$ at a given temperature is:

\begin{equation}
\frac{d n}{d t} = -n^2 \nu
\label{clusterdecay}
\end{equation}
Using integration by parts we solve this differential equation for the boundary of the cluster density $n_1$ before, and $n_2$ after an annealing step of a fixed time interval $\Delta t$ resulting in
\begin{equation}
\nu(T) = \frac{1}{\Delta t}\left(\frac{1}{n_2} - \frac{1}{n_1}\right).
\label{hoppingrate}
\end{equation}

We emphasize here that our crude approximation effectively averages over different size dependent interaction frequencies $\nu$ for a given size distribution. However, using the annealing time $\Delta t$ and the annealing temperature $T$, this appoach allows one to derive the temperature dependence of $\nu$ from an annealing sequence as shown in \ref{anneal2} (a)-(f). 

The resulting Arrhenius plots are shown in Fig. \ref{dynamics} (h). Activation energies are between 0.38\,eV and 0.75\,eV.
The resulting $\nu_0$ lie between 1.4\,Hz and 500\,Hz.
These attempt frequencies are much lower than a typical phonon frequency and also much lower than what is found for the diffusion of adatoms and small adclusters (compare e.g. \cite{wang1990}). 
According to transition state theory the low $\nu_0$ point to exceptionally large differences of the partition function of clusters in the bound state versus clusters in the transition state. 
The diffusion parameters have to be viewed as \emph{effective diffusion parameters} for an ensemble of clusters comprising all size effects and not as the properties of an individual cluster. This is also illustrated by the large variations in parameters for the observed cluster lattices from Pt with different average cluster size.

\begin{table}
\begin{center}
\begin{tabular}{|c||c|c||c|c|c|}
\hline
Clusters 	  & $E_a$ 	& $\Delta E_a$ 	& $\nu_0$	& $\Delta \nu_{0,-}$ & $\Delta \nu_{0,+}$ \\
\hline
Ir, 0.45\,ML (I)  &  0.41\,eV 	& 0.02\,eV	& 1.4\,Hz 	& 0.5\,Hz	& 0.8\,Hz \\
Ir, 0.45\,ML (II) &  0.75\,eV	& 0.2\,eV	&  67\,Hz	&  65\,Hz	& 2700\,Hz \\
Ir, 0.45\,ML 	  &  0.28\,eV   & 0.08\,eV	& 0.06\,Hz	& 0.03\,Hz	& 0.05\,Hz \\
Pt, 0.25\,ML	  &  0.60\,eV   & 0.08\,eV	& 500\,Hz	& 430\,Hz	& 3100\,Hz \\
Pt, 0.70\,ML	  &  0.38\,eV	& 0.02\,eV	& 6.2\,Hz	& 2.3\,Hz	& 3.7\,Hz \\
W,  0.44\,ML	  &  0.47\,eV	& 0.04\,eV	&  33\,Hz	&  20\,Hz	& 52\,Hz \\
\hline
\end{tabular}
\end{center}
\caption{Diffusion parameters and corresponding statistical errors as derived from $\nu(T)$, for the cases of Ir, Pt an W. The parameters have as well been determined separately for two sections (I) and (II) of the dataset as indicated in Fig.  \ref{anneal2} (h)
Attempt frequencies have an statistical asymmetric error, so that negative and positive one are given. Systematic errors may be larger (see text).}
\label{tab:diffusionparams}
\end{table}
For the case of Ir as shown in Fig. \ref{anneal2} (g) and (h), there is a distinct discontinuity in the cluster density and consequently in the estimated interaction frequencies $\nu$ between 550\,K and 650\,K. Interestingly this transition coincides with all single layered clusters dying out.
We interpret the discontinuity as a result of different diffusion parameters for single layered and multi layered clusters.

\begin{figure}
\begin{center} 
\includegraphics[width=0.6 \linewidth]{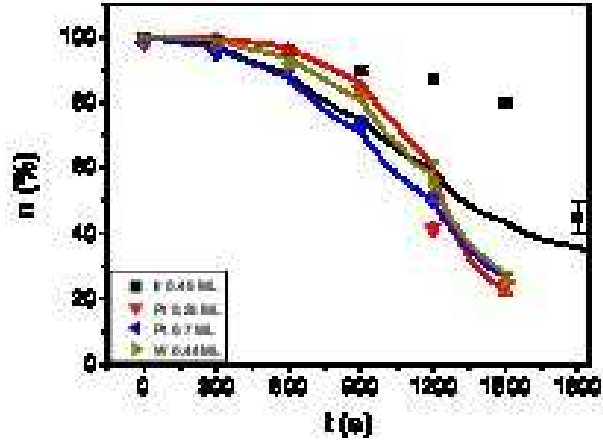}
\end{center}
\caption{Cluster density $n$ evolution with a Monte Carlo simulation using average diffusion parameters. The temperature is increased every 300\,s corresponding to Fig. \ref{anneal2}(g) and (h).
Cluster densities after the annealing steps as in figure \ref{anneal2} are reshown for comparison.}
\label{simulation}
\end{figure} 
This approach of effective diffusion parameters is checked for consistency with a kinetic Monte Carlo simulation.
The algorithm for the simulation is based on work by Bortz at al. \cite{bortz1975}.
The cluster lattice is modeled as a hexagonal lattice with clusters, which can hop to adjacent sites with a frequency $\nu = \nu_{\rm 0} e^{E_{\rm a}/k_{\rm B} T}$ based on  $\nu_{\rm 0}$ and $E_{\rm a}$ as given in table \ref{tab:diffusionparams}.
The cluster lattice dwells at each annealing temperature for 300\,s and is heated successively to higher temperature annealing intervals. Time for cooling, imaging, and reheating is omitted.
The simulation reproduces the cluster densities well as shown in figure \ref{simulation}. Kinks occur in the curve, where the temperature changes. 
Not surprising, Iridium is an exception to the good match because one kind of clusters dies out rapidly and the approximation of effective diffusion parameters breaks down for the examined temperature range.

\subsection{Binding sites of clusters in the superlattice}

\begin{figure}
\begin{center} 
\includegraphics[width=0.6 \linewidth]{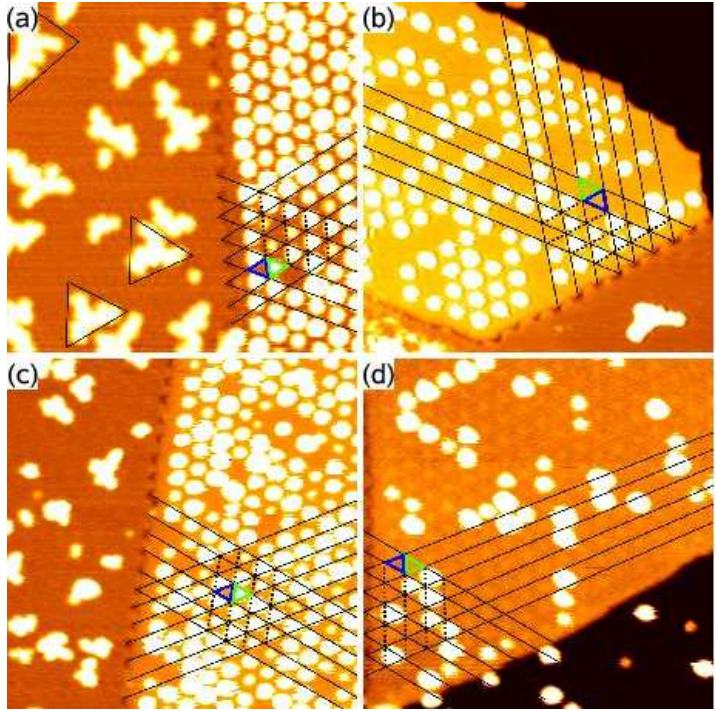}
\end{center}
\caption{Binding site determination for (a) Ir, (b) Pt, (c) W and (d) Re clusters. Deposition was performed at 300\,K and (a) $\Theta = 0.10$\,ML, $s_{\rm{av}} = 9$\,atoms; (b) $\Theta = 0.04$\,ML, $s_{\rm{av}} = 6$\,atoms; (c) $\Theta = 0.04$\,ML, $s_{\rm{av}} = 4$\,atoms; (d) $\Theta = 0.03$\,ML, $s_{\rm{av}} = 10$\,atoms.
Due to the moiré the graphene flakes form a jagged edge if in contact with a $\left\langle 1\bar{1}0\right\rangle$/$\left\{100\right\}$ microfacet or A-step. The dark tips of the jagged graphene edge are atop site areas. Fixing the grid of moiré unit cells to these positions enables a binding site assignment. The clusters always sit in the green triangular half-unit cells pointing away from the step (see text).}
\label{binding}
\end{figure} 

Experimentally \cite{ndiaye2006,ndiaye2008} and by calculations \cite{feibelman2008} we find Ir clusters to adsorb preferentially in hcp regions of the moiré unit cells. These sites differ from the fcc regions only by the fact that instead of a threfold coordinated fcc hollow site a threefold coordinated hcp hollow site is centered in the carbon ring. They differ significantly from the atop-type area which has an atop site centered in the carbon ring. At low growth temperature we find for Ir deposition also fcc regions to be populated by small clusters, consistent with the similarity of the two areas \cite{ndiaye2008}. It is not at all evident that also other materials adsorb preferentially to hcp regions. Specifically for materials with a different crystal structure like W with its bcc structure or Re with the hcp crystal structure we would not be surprised to find the clusters adsorbed preferentially to fcc regions or to be even unspecific to the small difference caused by the second layer underneath the Ir surface. To obtain cluster binding sites experimentally we make use of the fact that the graphene sheets on Ir form a jagged zig-zag edge when in contact with a $\left\langle 1\bar{1}0\right\rangle$/$\left\{100\right\}$ microfacet or A-step of the substrate. The step undulation of the graphene sheet has the moiré periodicity. The protrusions of the graphene flake's edge are bowing out towards the Ir terrace at atop-type areas. This fact does evidently not depend on the deposited material and allows us unambiguous cluster adsorption site assignment. In Fig. \ref{binding} for Ir, Pt, W and Re the corners of the moiré unit cell grids are fixed to these atop-type areas. We find that the clusters are always located in the triangular half-unit cells pointing away from the one dimensional graphene-Ir interface (green triangles in Fig. \ref{binding}). According to our unit cell assignment (compare Fig. 1 of reference \cite{ndiaye2006}) these are hcp regions. 

Let us point out that in Fig. 8 of reference \cite{ndiaye2008} and in the accompanying text of Sec. 8 the labeling and use of hcp region and fcc region is erroneously interchanged. 

\subsection{Towards cluster superlattice materials}

\begin{figure}
\begin{center} 
\includegraphics{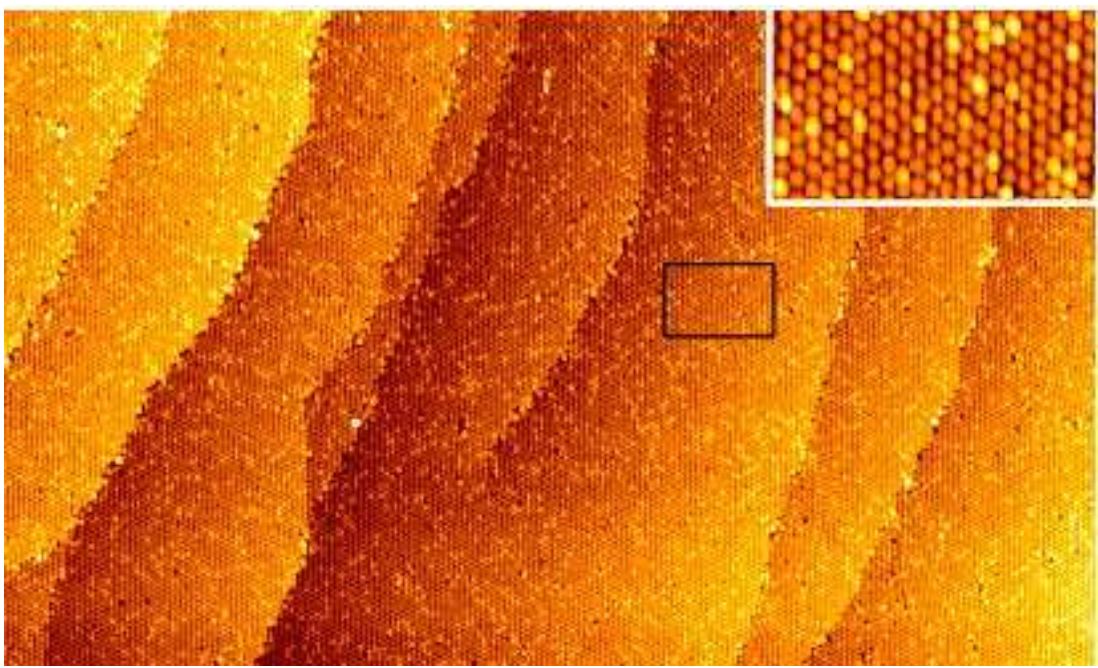}
\end{center}
\caption{Ir cluster superlattice grown at 300\,K with $\Theta = 0.80$\,ML resulting in $s_{\rm{av}} = 70$ on graphene prepared by temperature programmed growth followed by chemical vapor deposition and extending over the entire sample. Image size $0.5\,\mu$m~$\times~0.3\,\mu$m, inset 500\,\AA~$\times$~300\,\AA.}
\label{CVDcluster}
\end{figure}

To probe the properties of cluster superlattices by averaging techniques and to investigate their suitability for potential applications it is necessary to cover a sample macroscopically with a cluster superlattice. This need is evident if one considers e.g. the analysis of reaction products from a cluster superlattice in nanocatalysis. The presence of the bare metal would result in additional peaks in thermal desorption spectra and certainly complicate the data analysis. To establish a macroscopic cluster superlattice the graphene moiré must cover the metal substrate entirely, be of unique orientation with an as large as possible moiré supercell which displays upon deposition of suitable materials local rehybridization. While it is likely that a number of graphene moirés on different metals fulfill all these conditions, so far they have been proven only for graphene moirés on Ir(111). Some optimization of the graphene growth procedure was necessary to achieve simultaneously full graphene coverage and a single orientation of the graphene and the graphene moiré \cite{coraux2009,coraux2008,vangastel2009}. Fig. \ref{CVDcluster} displays a large scale STM topograph visualizing to a certain extent the quality of the available substrate. The cluster superlattice is present in the entire topograph in unique orientation and even steps merely present locations where a line of clusters is missing, but without disturbing the overall alignment of the superlattice. 

\section{Conclusion}

In conclusion, we have established that the graphene moiré on Ir(111) is a versatile and active template for cluster superlattice growth of a great variety of materials and with macroscopic lateral extension. If necessary, techniques like low temperature growth or cluster seeding may permit cluster superlattice growth for cases, where simple room temperature deposition fails. 
The high thermal stability of the cluster superlattices and the ability to grow them on macroscopic areas opens new opportunities for fundamental cluster research and applications.

\section{Acknowledgements}
Work at Cologne University was supported by DFG through the project ``Two Dimensional Cluster Lattices on Graphene Moirés''. T.M. acknowledges useful discussion with Peter J. Feibelman.
\vspace{2 cm} 
 
\section*{Bibliography} 
\bibliographystyle{unsrt}
\bibliography{cluster_universal}

\begin{thebibliography}{10}
\bibitem{footnote}
In fact for a flat 19 atom Ir cluster adsorbed in an hcp-type area on the
  graphene moiré we find an average Ir-Ir nearest neighbour separation of
  2.54\,\AA.
\end{thebibliography}

\end{document}